\documentclass[sigconf,10pt]{acmart}
\renewcommand\footnotetextcopyrightpermission[1]{} 
\usepackage{booktabs} 
\usepackage{microtype}
\usepackage[utf8]{inputenc}

\newcounter{informantcounter}
\newcommand{\informant}[1]{%
    \ifcsname #1@count\endcsname%
        X\csname #1@count\endcsname%
    \else%
        \stepcounter{informantcounter}%
        \expandafter\xdef\csname #1@count\endcsname{\theinformantcounter}%
        X\theinformantcounter%
    \fi%
}

\acmConference[FAT/ML 2017]{Fairness, Accountability and Transparency in Machine Learning}{2017}{Halifax, Canada}

\makeatletter
\def\blfootnote{\xdef\@thefnmark{}\@footnotetext}
\makeatother

\usepackage{csquotes}
\newcommand*{\myenquote}[1]{\enquote{{\itshape#1}}}
\usepackage{microtype}

\begin{document}
\title{Logics and practices of transparency and opacity in real-world applications of public sector machine learning}


\author{Michael Veale}
\orcid{0000-0002-2342-8785}
\affiliation{%
  \institution{Department of Science, Technology, Engineering \& Public Policy (STEaPP)\\ University College London}
  \streetaddress{Gower Street, London} 
}
\email{m.veale@ucl.ac.uk}

\sloppy

\begin{abstract}
	Machine learning systems are increasingly used to support public sector decision-making across a variety of sectors. Given concerns around accountability in these domains, and amidst accusations of intentional or unintentional bias, there have been increased calls for transparency of these technologies. Few, however, have considered how logics and practices concerning transparency have been understood by those involved in the machine learning systems already being piloted and deployed in public bodies today. This short paper distils insights about transparency on the ground from interviews with 27 such actors, largely public servants and relevant contractors, across 5 OECD countries. Considering transparency and opacity in relation to trust and buy-in, better decision-making, and the avoidance of gaming, it seeks to provide useful insights for those hoping to develop socio-technical approaches to transparency that might be useful to practitioners on-the-ground.
\end{abstract}

\maketitle
\blfootnote{\textit{4th Workshop on Fairness, Accountability, and Transparency in \mbox{Machine} Learning (FAT/ML 2017), Halifax, Canada, August 14, 2017.}}

\fbox{\parbox{\linewidth}{\textbf{\textit{Note to readers: an since-published, extended version of this paper can be found at:} Veale M., Van Kleek M., \& Binns R. (2018). `Fairness and accountability design needs for algorithmic support in high-stakes public sector decision-making' Proceedings of the 2018 CHI Conference on Human Factors in Computing Systems (CHI’18), \url{http://doi.org/10.1145/3173574.3174014}. \\ \textit{Please refer to that one in place of this one.}}}}

\section{Introduction}
A range of machine learning (ML) systems are being developed, piloted and deployed in public sector organisations across the world. 27 actors involved in public sector ML projects within five OECD countries were interviewed about their experiences building and deploying machine learning systems for decision-support, and how they related to a variety of public sector values often considered broadly held or aimed for (see \cite{jorgensen2007public}). These projects were ML-based decision support projects in sectors including child protection, policing, fire response, taxation, justice and the prison system, interior affairs and emergency helicopter support. This paper presents and summarises responses relating to just one of several salient themes that emerged from these interviews --- how individuals and organisations have navigated practical challenges and choices around transparency and opacity. The paper does not seek to be a representative survey; the aim instead is to illustrate the minimum level of diversity in this area, and present opinions and experiences rarely discussed today within formal communities considering algorithmic transparency or accountability.

\section{Method}
27 individuals agreed to be interviewed in late 2016, predominantly public servants and attached contractors either in modelling or project management. Informants were identified with an \emph{ad hoc} sampling approach due to dispersion of practice, and that no identified public body appears to maintain a reliable central register of projects. Projects and contacts were assembled from grey literature, freedom of information requests, snowball sampling, direct inquiries with organisation contacts, and the use of news databases including \textit{Factiva} and \textit{LexisNexis}. Conversations were not recorded for reasons of sensitivity, and instead all notes were transcribed into long form immediately after the interview terminated. Some terminology was amended to increase the difficulty of project re-identification. Just over one-fifth of the informants were female. The U.S. was not one of the countries containing examined projects, and the 5 countries from which examples \textit{are} drawn are spread over three continents. To organise content, interviews were coded according to a common public sector values framework \cite{jorgensen2007public} using \textit{NVivo 11 for Mac}.

\section{Findings and discussion}

Transparency is often seen as desirable in high-stakes machine learning systems, and much attention has been directed to its promotion. Many recent voices have highlighted the layered nature of transparency \cite{Burrell2016} or its limits as a policy approach \cite{edwardsveale}, and work in the HCI and computer science communities over the last two decades on developing a range of explanation facilities for machine learning has seen renewed vigour. Yet practitioners are rarely asked about their opinions, experiences and strategies in this area. Three main strands concerning transparency and opacity are highlighted here: \textit{transparency for organisational trust and buy-in}, \textit{transparency for better decision-making} and \textit{opacity to mitigate internal and external gaming}.

\subsection{Transparency for trust and buy-in}

Several interviewed practitioners noted that organisational pressure led them to make more `transparent' machine learning systems. Detection systems for fraudulent tax returns illustrate this. The analytics lead at one tax agency [\informant{taxlead1}] noted that they \myenquote{have better buy-in} when they provide the logic of their machine learning systems to internal customers, whilst their counterpart in another tax agency [\informant{taxlead2}] described a need to \myenquote{explain what was done to the business user}. Both these individuals and the modellers around them emphasised they \textit{had} in-house capability for more complex machine learning systems, such as support vector machines or neural networks, but often chose against them for these reasons. Instead, many of the systems that ended up being deployed were logistic regression or random forest--based.

Some saw transparency in relation to input variables more than model family. One contractor that constructed a random-forest based risk score for gang members around knife crime on behalf of a global city's police department [\informant{police-c1}] described an \myenquote{Occam's razor} process, where they started with 18,000 variables, working down to 200, then 20, then 8 --- \myenquote{because it's important to see how it works, we believe}. To steer this, they established a target percentage of accuracy with the police department \textit{before} modelling --- around 75\% --- which they argued helped them avoid trading off transparency. When users of analytics are not \myenquote{confident they know what a model is doing}, they \myenquote{get wary of picking up protected characteristics}, noted the modelling lead at a tax agency [\informant{taxmodel1}]. To make this more transparent, the police contractor above [\informant{police-c1}] would \myenquote{make a model with and without the sensitive variables and see what lift you get in comparison}, presenting those options to the client to decide what was appropriate.

Over and under-reliance on decision support, extensively highlighted in the literature on \textit{automation bias} \cite{Skitka:1999il,Dzindolet:2003bl}, were highlighted in relation to transparency. A lead machine learning modeller in a national justice ministry [\informant{justice1}], allocating resources such as courses within prisons with machine learning, described how linking systems with professional judgement \myenquote{can also mean that it only is used when it aligns with the intuition of the user of the system}. To this end some informants considered how to bring discretion into decision-support design. A lead of a geospatial predictive policing project in a major city [\informant{police1}] noted that they designed a user interface

\begin{displayquote} \itshape
to actively hedge against [officers resenting being told what to do by models] by letting them look at the predictions and use their own intuition. They might see the top 3 and think `I think the third is the most likely' and that's okay, that's good. We want to give them options and empower them to review them, the uptake will hopefully then be better than when us propellorheads and academics tell them what to do...
\end{displayquote}

Transparency in outputs for discretion had different effects in different application areas. The former lead of a national predictive policing strategy [\informant{police-b2}] explained

\begin{displayquote} \itshape
We [use machine learning to] give guidance to helicopter pilots, best position them to to optimise revenue --- which means they need to follow directions. They lose a lot of flexibility, which made them reluctant to use this system, as they're used to deciding themselves whether to go left or right, not to be told `go left'! But it's different every time. There were cases where agents were happy to follow directions. Our police on motorcycles provide an example of this. They were presented with sequential high risk areas where criminals should be and would go and apprehend one after another --- and said ``yes, this is why we joined, this is what we like to be doing!'' The helicopters on the other hand did not like this as much.	
\end{displayquote}

This raised issues around the `ethics of following lists'. The analytics lead at one regional police department, designing a range of in-house machine learning models for both geospatial prediction and individual offender/victim prediction [\informant{police-a1}], sought advice from their internal ethics committee on how to use the prioritised lists given as outputs

\begin{displayquote} \itshape
we had guidance from the ethics committee on this. We were to work down the list, allocating resources in that order, and that's the way they told us would be the most ethical way to use them... It's also important to make clear that the professional judgement always overrides the system. It is just another tool that they can use to help them come to decisions.
\end{displayquote}

\subsection{Transparency for decision-making}

While the above informants focussed on transparency for buy-in, others emphasised how they sought practices of transparency to improve the decisions that were being made. One lead tax analyst [\informant{taxlead2}] described how transparency provided \myenquote{value-add, particularly where an administrative decision needs explaining to a customer, or goes to tribunal.} A lead of a national geospatial predictive policing project [\informant{police-b1}] discussed transparency in more social terms how the intelligence officers, who used to spend their time making patrol maps, now spent their time augmenting them.

\begin{displayquote} \itshape
We ask local intelligence officers, the people who read all the local news, reports made and other sources of information, to look at the regions of the [predictive project name] maps which have high predictions of crimes.  They might say they know something about the offender for a string of burglaries, or that building is no longer at such high risk of burglary because they local government just arranged all the locks to be changed. [...] We also have weekly meeting with all the officers, leadership, management, patrol and so on, with the intelligence officers at the core. There, he or she presents what they think is going on, and what should be done about it. 
\end{displayquote}

A modeller and software developer in the same project [\informant{police-b3}] emphasised the challenges in scaling up these social practices, as they were not as mobile as the software itself.

\begin{displayquote} \itshape
If you want to roll out to more precincts, they have to actually invest in the working process to transform the models into police patrols. To get more complete deployment advice... it takes a lot of  effort to get people to do that. What you see is that other precincts usually --- well, sometimes --- set up some process but sometimes it is too pragmatic. What I mean by this is that the role of those looking at the maps before passing them to the planner might be fulfilled by someone not quite qualified enough to do that.
\end{displayquote}

Similar sentiments were also echoed by individuals in national tax offices, particularly around the `trading' of models by large vendors.

Transparency within organisations also arose linked to the primary collectors of training data. One in-house modeller in a regional police department [\informant{police-a2}], building several machine learning models including one to predict human trafficking hotspots, described how without better communication of the ways the models deployed worked, they risked large failure.

\begin{displayquote} \itshape
Thankfully we barely have any reports of human trafficking. But someone at intel got a tip-off and looked into cases at car washes, because we hadn't really investigated those much. But now when we try to model human trafficking we only see human trafficking being predicted at car washes, which suddenly seem very high risk. So because of increased intel we've essentially produced models that tell us where car washes are. This kind of loop is hard to explain to those higher up.
\end{displayquote}

Cases like this blur transparency with communication. If machine learning systems become more important in public organisations, understanding the provenance of the data being used --- and establishing communications up and down that chain of provenance to understand when logics of collection or cleaning change --- becomes important.

\subsection{Opacity to avoid gaming}

On this topic, other organisations expressed concern that extensive transparency might encourage system gaming, both from within, by those using the decision support, and by the subjects and potential subjects affected by decisions being made. External gaming is often heard stated as a reason for opacity. One contractor developing predictive policing software for a world city [\informant{police-c1}] noted that concerns in his sector concerned \myenquote{criminal gangs that might send nine guinea pigs through the application process looking for loopholes to get arrested, just to find a tenth that highlights a way they can reliably get passports from under the noses of the authorities.} An analyst from a large NGO working in collaboration with the police on developing a predictive system to detect child abuse [\informant{ngo1}] noted that \myenquote{it's much harder to game when you've linked up lots of different aspects, education and the like}, although their colleague [\informant{ngo2}] warned that they were concerned about many sophisticated practices commonly used to game child protection systems also being challenging for ML-supported systems, such as \myenquote{turning professionals against each other} or the \myenquote{strategic withholding of consent at opportune moments.} The analytics lead at one tax agency [\informant{taxlead1}] explained that while they would publicly share the areas they were interested in modelling tax fraud for, such as sector or size, they were \myenquote{primarily concerned that if the model weights were public, their usefulness might diminish}.

Yet concerns around gaming do not end with the external decision subjects. Internal gaming, extensively highlighted in the public administration literature in relation to targets and New Public Management \cite{Bevan:2006vk}, was on the minds of many informants. One tax analytics lead [\informant{taxlead2}] worried that releasing the input variables and their weightings in a model could make their auditors investigate according to their perception of the model structure, rather than the actual model outputs --- where fairness analysis and bias\footnote{Many informants also explained their approaches to mitigating unfairness or discrimination in machine learning systems, which will be covered in a future paper.} could ostensibly be `controlled'. 

\begin{displayquote} \itshape
To explain these models we talk about the target parameter and the population, rather than the explanation of individuals. The target parameter is what we are trying to find --- the development of debts, bankruptcy in six months. The target population is what we are looking for: for example, businesses with minor problems. We only give the auditors [these], not an individual risk profile or risk indicators [...] in case they investigate according to them.
\end{displayquote}

There was concern about gaming within tax organisations, as well as outside it. Some tax auditors are tasked with using the decision-support from machine learning systems to inform their fraud investigations. Yet at the same time, the fraud they discover feeds future modelling; they are both decision arbiter and data collector. The effect these conflicting incentives might have on a model were highlighted by a different tax agency [\informant{taxlead2}]. When auditors accumulate their own wages, they noted that \myenquote{[i]f [an auditor] found an initial [case of fraud],  [the auditor] might want to wait for [the uncovered] individuals to accumulate it, which would create perverse incentives for action}.

\section{Concluding remarks}

This paper sought to provide a short overview of the logics and practices of transparency in the public sector. In particular, it pointed to some logics for transparency and opacity that go beyond accountability, concerning organisational buy-in, better decision-making, and avoiding both internal and external gaming. In the furore around algorithmic accountability, the importance of effective transparency within contexts of use has often been overlooked. Understanding these, and other institutional logics, will be key to both co-designing appropriate social and technical approaches for useful transparency, and seeing them implemented in important, real-world contexts. Three points are highlighted in conclusion.

Firstly, techniques to improve transparency on the ground are often complex, social, and not always apparent from `reverse engineering' approaches. Public sector actors have built up considerable social routines to augment or alter, or deliver model outputs in response to challenges found around the data and the model users. Broad preconceptions that individuals from system design to management blindly follow data and algorithms are often not just unjustified, but likely to patronise (and damage potential collaborations) with those deeply involved in building and maintaining such routines. Yet we also know very little about these practices: how they are, or are not transmitted with the software being implemented, how they develop and evolve, and the tangible impact they have on identifying or mitigating some of the issues of fairness that have been previously highlighted. As far back as 2007, academics have been working with the public sector in these specific areas, undertaking process evaluations of police use of machine learning for predictive mapping to understand these phenomena \cite{johnson2007}, but lately this approach seems to have fallen out of vogue in favour of attempting to reverse engineer systems, rather than engage in collaboration to understand the way technologies are perceived, used and evolved. This is useful, but limited. Transparency is a process that has both internally and externally-facing components, and insofar as these are both important and linked, they require rigorous academic investigation.

Secondly, while gaming is often suggested as a large barrier to transparency, internal gaming is much less considered than external gaming. Neither have a good evidence base in relation to decision-support systems, although the use of discretion around decision support has been studied in institutional contexts \cite{Bovens:2002jv,Jorna:2007wl,buffat2015street}. The majority of the concern in this area expressed by interviewees was pre-emptive, rather than based on experience in deployed machine learning. If true, then internal gaming is of particular concern, as individuals using the systems also often collect fresh data --- potentially challenging some basic statistical assumptions of stationarity that underpin both machine learning systems in general and approaches for fairness and `debiasing'.

Lastly, transparency has often been highlighted as being entwined with trade secrets and/or proprietary systems, yet this was not an issue highlighted by informants. Many of the public sector machine learning systems found here were developed in-house or bespoke, in collaboration with academia, or by paying consultants for time and retaining the IP. Recidivism systems in the UK and Dutch prison systems openly publish the model weights for their logistic regression systems \cite{mooreOASys,tollenaar2016}, while Durham Police have openly presented the variable importance scores in their recidivism-anticipating random forests \cite{urwin_2017}. In general not only were interviewees open and generous with access to their software, but the majority interviewed were well aware of many of the technical challenges highlighted by research communities concerned about machine learning and ethics. Some were even aware of specific pieces of literature in this area --- approaches that they would have liked to test and deploy, were they available in commonly available, vetted software packages in open-source programming languages. Transparency might be useful for journalists, or for auditors, but it is also useful within organisations. Where possible, starting with the needs of machine learning users, working in a transdisciplinary manner, might often be more fruitful than only pressuring these organisations from outside.

\section*{Acknowledgements}
This work is supported by the Engineering and Physical Sciences Research Council (EPSRC) grant EP/M507970/1. The author acknowledges further funding from the World Bank GFDRR and a grant from University College London's STEaPP pump-priming fund. The project was approved by UCL's Research Ethics Committee, reference 7617/001. Additional thanks go to three anonymous reviewers who made comments which improved this piece.

\bibliographystyle{ACM-Reference-Format}
\bibliography{fatml_bibliography.bib} 

\end{document}